\documentclass[final]{aipproc}

\layoutstyle{6x9}

\begin{document}

\title{Scalar meson mass from renormalized One Boson Exchange Potentials\footnote{Talk presented by ACC at
{\em SCADRON 70 Workshop on ``Scalar Mesons and Related Topics''}, Lisbon, 11-16 February 2008}}


\author{A. Calle Cord\'on}{
  address={Departamento de F\'isica At\'omica, Molecular y Nuclear,\\
  Universidad de Granada,\\
  E-18071 Granada, Spain.}}

\author{E. Ruiz Arriola}{
  altaddress={Departamento de F\'isica At\'omica, Molecular y Nuclear,\\
  Universidad de Granada,\\
  E-18071 Granada, Spain.}}

\begin{abstract}
We determine the mass and strength of the scalar meson from $NN$
scattering data by renormalizing the One Boson Exchange Potential.
This procedure provides a great insensitivity to the unknown short
distance interaction making the vector mesons marginally important
allowing for SU(3) couplings in the $^1S_0$ channel.  The scalar meson
parameters are tightly constrained by low energy np. We discuss
whether this scalar should be compared to the recent findings based on
the Roy equations analysis of $\pi\pi$ scattering.
\end{abstract}

\maketitle


\section{Introduction}
  
Half a century ago Johnson and Teller~\cite{PhysRev.98.783} suggested
the need for a scalar-isoscalar meson with a mass $ \sim 500{\rm MeV}
$ to provide saturation and binding in nuclei. In a way this was the
starting point for One-Boson-Exchange (OBE) Potentials where, in
addition to the pion, all possible resonances would be
included~\cite{Partovi:1969wd,Machleidt:1987hj,Machleidt:2000ge}.
Despite their undeniable success describing NN scattering data, there
has always been some arbitrariness on the scalar meson mass and
coupling constant to the nucleon, partly stimulated by a lack of other
sources of information, definitely helping the fits. The relation of
the ubiquitous scalar meson in nuclear physics and NN forces in terms of
correlated two pion exchange has been pointed out many
times~\cite{Partovi:1969wd,Machleidt:1987hj} (see
e.g. \cite{Kaiser:1998wa,Oset:2000gn,Donoghue:2006rg} for a discussion
in a chiral context).

The quest for the existence of the $0^{++}$ resonance (commonly
denoted by $\sigma$) has finally culminated with its inclusion in the
PDG~\cite{Yao:2006px} as the $f_0 (600)$ seen as a $\pi\pi$ resonance,
where a spread of values ranging from $400-1200 {\rm MeV}$ for the
mass and a $600-1200 {\rm MeV}$ for the width are
displayed~\cite{vanBeveren:2002mc}. The uncertainties have recently
been sharpened by a determination based on Roy equations and chiral
symmetry~\cite{Caprini:2005zr} yielding the value $m_\sigma - i
\Gamma_\sigma /2 = 441^{+16}_{-8} - i 272^{+9}_{-12} {\rm MeV}$; the
lowest resonance in the hadronic spectrum. It is mandatory and perhaps
possible to scrutinize its role in hadronic phenomenology all
over. Here, we approach the problem from NN scattering in the $^1S_0$
channel from a renormalization viewpoint as applied to the OBE
potential (without explicit inclusion of $2\pi$ exchange) and try to
see the connection to $\pi\pi$ scattering.

\section{The traditional approach to OBE potentials}

The field theoretical OBE model of the NN
interaction~\cite{Machleidt:1987hj} includes all mesons with masses
below the nucleon mass, i.e., $\pi$, $\eta$, $\rho(770)$ and
$\omega(782)$, in addition with a scalar-isoscalar boson.  Dropping
$\eta$ and $\rho$ because of their small couplings, the $^1 S_0$ NN
potential is
\begin{eqnarray}
V(r) = -\frac{ g_{\pi NN}^2 m_{\pi}^2} {16 \pi M_N^2}
\frac{e^{-m_{\pi} r}} {r} - \frac{ g_{\sigma NN} ^2}{4 \pi}\frac{e^{-
m_{\sigma} r}} {r} + \frac{g_{\omega NN}^2}{4 \pi}\frac{e^{-m_{\omega}
r}}{r} + \cdots
\label{eq:pot}
\end{eqnarray}
As is well known~\cite{Partovi:1969wd} any perturbative determination
of a potential suffers from off-shell ambiguities (even in the Born
approximation), particularly because of relativistic finite mass
corrections which may be smoothly shifted between entirely energy
dependent and local potentials or energy independent and nonlocal
potentials. This trading between retardation and nonlocality may
become sizable at short distances scales, where the interaction is
unknown anyhow, and the particular choice is completely arbitrary.
Our renormalization scheme will be such that, as suggested by Partovi
and Lomon we ignore both retardation and nonlocality in the long
distance limit~\cite{Partovi:1969wd}, as well as the exponentially $
\sim e^{-2 M_N r } $ suppressed $N \bar N $ cut. Relativistic effects
are only kept by renormalization of the couplings, but the effect is
small~\footnote{This corresponds to $ g_{\sigma NN}^2 \to g_{\sigma
NN}^2 / \sqrt{1-m_\sigma^2 / 4 M_N^2}$ and $ g_{\omega NN}^2 \to
g_{\omega NN}^2 / \sqrt{1-m_\omega^2 / 4 M_N^2}$ }.


In any case, one should bear in mind that NN scattering in the elastic
region below pion production threshold involves CM momenta $p <
p_{max} = 400$ MeV. Given the fact that $1/m_{\omega} = 0.25
\mathrm{fm} \ll 1/p_{max} = 0.5\mathrm{fm}$ we expect heavier mesons
to be irrelevant, and $\omega$ itself to be marginally important, even
in s-waves, which are most sensitive to short distances. In order to
illustrate this, we take $m_{\pi} = 138$MeV, $M_{N} = 939$MeV,
$m_{\omega} = 783$MeV and $g_{\pi NN} = 13.1$, which seem firmly
established, and treat $m_{\sigma}$, $g_{\sigma NN }$ and $g_{\omega
NN}$, as fitting parameters. As we show now, this vector meson
irrelevance has not been fulfilled in the conventional approach to NN
scattering, forcing too large $g_{\omega NN}$ couplings. Actually, in
the standard approach the scattering phase-shift $\delta_0(p)$ is
computed by solving the (s-wave) Schr\"odinger equation
\textbf{r}-space
\begin{eqnarray}
-u''_p(r) + M_{N}\,V(r)\,u_p(r) &=& p^2\,u_p(r) \label{eq:Scrod-p} \\ 
u_p(r) &\to& \frac{\sin{\left(p r + \delta_0 (p)\right)}}{\sin{\delta_0(p)}}
\label{eq:up-asymp}
\end{eqnarray}
with a regular boundary condition at the origin
$u_p(0)=0$\footnote{This boundary condition obviously implies a
knowledge of the potential in the whole interaction region, and it is
equivalent to solve the Lippmann-Schwinger equation in
\textbf{p}-space.}.  Moreover, for a short range potential such as the
one in Eq.~\eqref{eq:pot} one also has the Effective Range Expansion
(ERE)
\begin{eqnarray}
p \cot{\delta_0 (p)} = -\frac{1}{\alpha_0} + \frac{1}{2}\,r_0\,p^2 +
v_2\,p^4 + \cdots
\label{eq:ere}
\end{eqnarray}
where the \textit{scattering length} $\alpha_0$ and the
\textit{effective range} $r_0$ are defined by the asymptotic behavior
of the zero energy wave function. In the usual
approach~\cite{Machleidt:1987hj,Machleidt:2000ge} everything is
obtained from the potential assumed to be valid for $0\leq r < \infty$
\footnote{In practice, strong form factors are included mimicking the
finite nucleon size and reducing the short distance repulsion of the
potential, but the regular boundary condition is always kept.}. In
addition, due to the \textit{unnaturally large} NN $^1S_0$ scattering
length ($\alpha_0 \sim -23 {\rm fm}$), any change in the potential $V
\to V + \Delta V$ has a dramatic effect on $\alpha_0$, since one 
obtains
\begin{eqnarray}
\Delta\alpha_0 = \alpha_0^2 M_N \int_0^{\infty} \Delta V(r)
u_0(r)^2\mathrm{d}r
\end{eqnarray}
and thus the potential parameters \textit{must be fine tuned}, and in
particular the short distance physics. A fit to the np data of
Ref.~\cite{Stoks:1994wp} yields two possible but incompatible
scenarios: $m_{\sigma }~=~477.0(5)$MeV, $g_{\sigma NN}~=~8.76(4)$,
$g_{\omega NN}~=~7.72(4)$ with $\chi^2/DOF~=~0.85$ and $m_{\sigma
}~=~556.34(4)$MeV, $g_{\sigma NN}~=~13.044(2)$, $g_{\omega
NN}~=~12.952(2)$ with $\chi^2/DOF~=~0.52$. The small errors should be
noted. The ambiguity in this solution is a typical inverse scattering
one; note that despite the $\omega$ being repulsive, the total
potential is not repulsive at short distances, and the corresponding
couplings and scalar mass are determined to high accuracy but
incompatible. This is just opposite to our expectations and we may
regard these fits, despite their success in describing the data, as
unnatural.

\section{Renormalization of the OBE potential}

To overcome the unphysical short distance sensitivity we implement the
renormalization viewpoint (see
e.g. Ref.~\cite{PavonValderrama:2005wv,PavonValderrama:2007nu}). In
the simplest version one proceeds as follows
\begin{itemize}
\item For a given $\alpha_0$ integrate in the zero energy wave
function $u_0 (r)$ down to the cut-off radius $r_c$. This is the
renormalization condition. 
\begin{eqnarray}
-u''_0(r) + M_{N}\,V(r)\,u_0(r) &=& 0\\
u_0(r) &\to& 1 -\frac{r}{\alpha_0}
\end{eqnarray}
\item Impose self-adjointness to get the finite energy wave function
$u_p (r_c)$, 
\begin{eqnarray}
u'_p(r_c)\,u_0(r_c) - u'_0(r_c)u_p(r_c) = 0
\end{eqnarray}
\item Integrate out the finite energy wave function $u_p(r)$,
Eq.~\eqref{eq:Scrod-p}, to determine the phase shift $\delta_0(p)$
from Eq.~\eqref{eq:up-asymp}.
\item Remove the cut-off $r_c\to 0$ to ensure \textit{model
independence}.
\end{itemize}

\begin{figure}
\includegraphics[height=.36\textheight,angle=270]{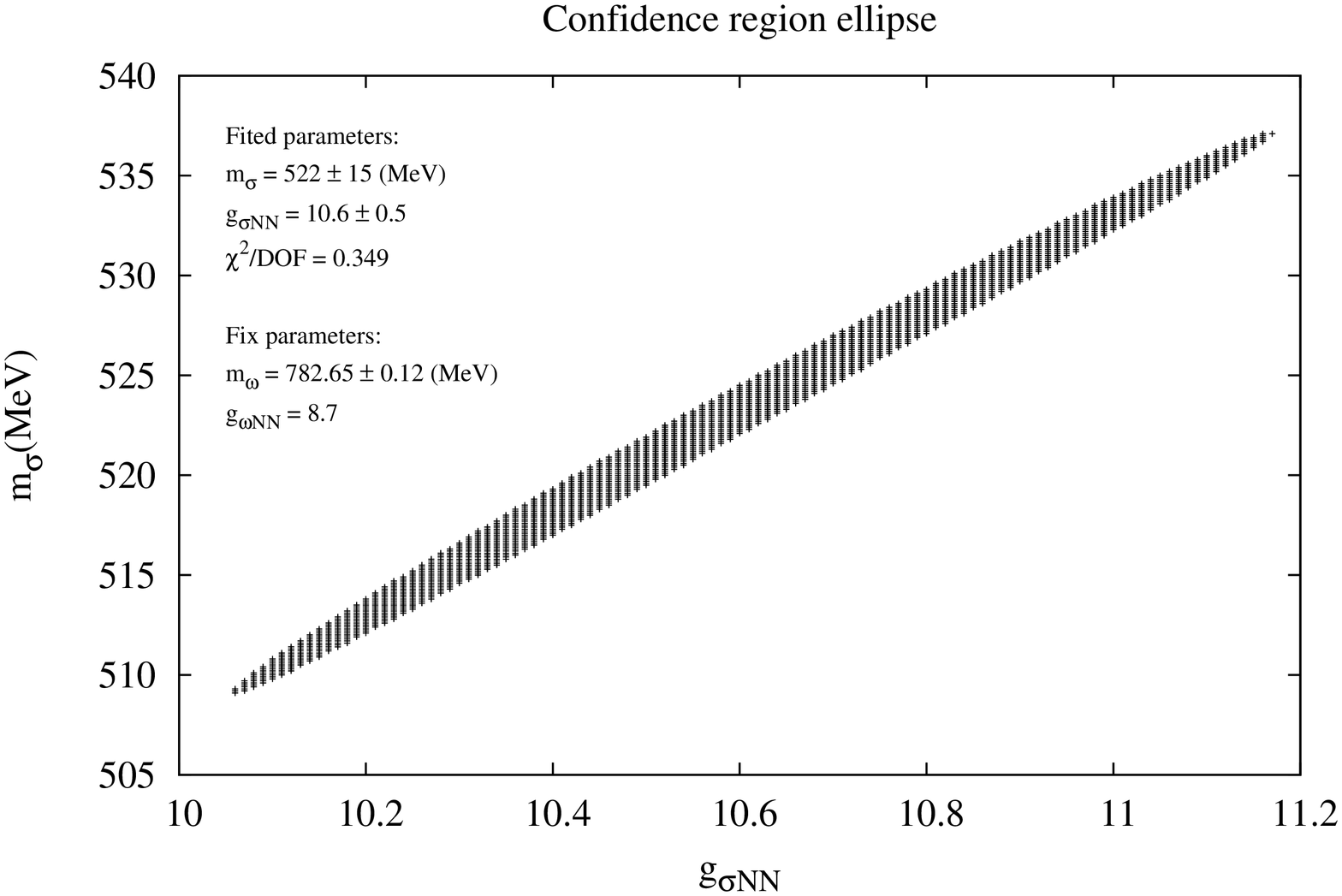}
\includegraphics[height=.36\textheight,angle=270]{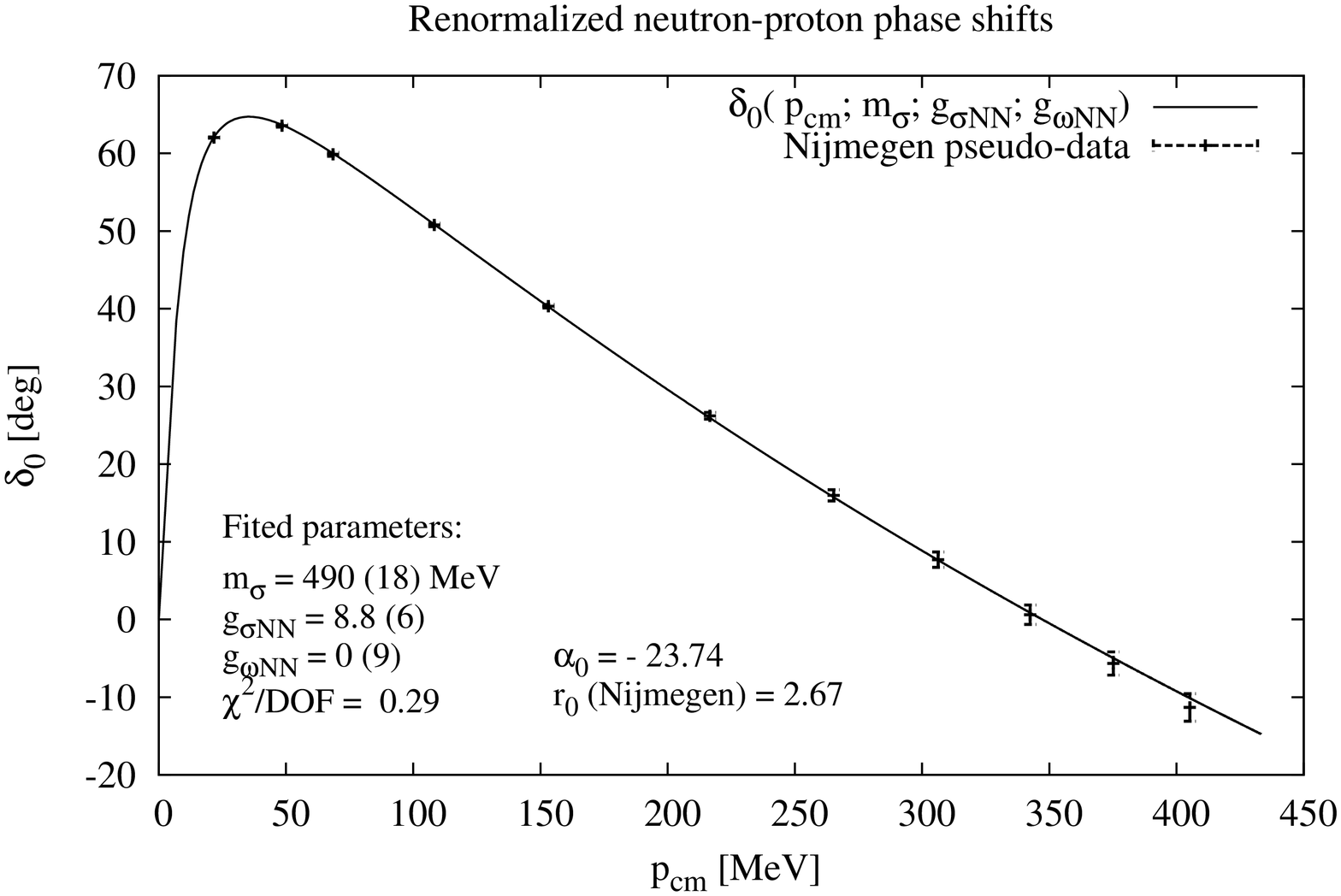}
\caption{ Left: $\Delta \chi^2 = 1 $ confidence level ellipse in the
$g_{\sigma NN}-m_\sigma$ plane for $g_{\omega NN}=8.7$. Right:
Renormalized OBE $^1S_0$ pn phase shifts (in degrees) as a function of
CM momentum. Data from \cite{Stoks:1994wp}.}
\label{fig:fits}
\end{figure}

This procedure allows us to compute $\delta_0(p)$ (and hence the next
order's parameters $r_0$,~$v_2$) from $V(r)$ and $\alpha_0$ as
independent information. Note that this is equivalent to consider, in
addition to the regular solution, the irregular one\footnote{In
momentum space this can be shown to be equivalent to introduce one
counterterm in the cut-off Lippmann-Schwinger equation, see
Ref.~\cite{Entem:2007jg} for a detailed discussion on this
connection.}. A fit of the potential Eq.~\eqref{eq:pot} to the np data
of Ref.~\cite{Stoks:1994wp} using the renormalization method gets
$m_{\sigma}~=~490(18)$MeV, $g_{\sigma NN}~=~8.8(6)$, $g_{\omega
NN}~=~0(9)$ with $\chi^2/DOF~=~0.29$. Note that $g_{\omega NN}$ is,
not only small but mostly irrelevant, so we consider this fit
natural. A consequence of this is that we could take the SU(3) value
$g_{\omega NN} = 3 g_{\rho NN} - g_{\phi NN} $ which on the basis of
the OZI rule, $g_{\phi NN} =0$, Sakurai's universality $ g_{\rho NN} =
g_{\rho \pi \pi} /2$ and the KSFR relation $ 2 g_{\rho \pi \pi}^2
f_\pi^2 = m_\rho^2 $ yields $g_{\omega NN} \sim 8.7 $ for which we get
$m_\sigma = 522(10) {\rm MeV}$, $g_{\sigma NN} = 10.5(5) $ and $\chi^2
/ DOF= 0.3$ with a strong linear correlation (see
Fig.~\ref{fig:fits}).

\section{What sigma ?}

Besides the numerical coincidence it is not obvious whether
or not are we entitled to identify the $NN$-scalar with the
$\pi\pi$-scalar because the $\pi\pi$-scalar has a large width, which
suggests that this state decouples~\footnote{A simple modification
such as $ V_\sigma (r) \to V_\sigma(r) \cos ( \Gamma_\sigma r/2 ) $
provides an inadmissible mid-range repulsion.}. We suggest a large
$N_c$ motivated scenario where this identification might actually
become compelling. The authors of \cite{Oset:2000gn} suggest that the
potential due to iterated $2 \pi $ scattering can be written for
non-vanishing distances
\begin{eqnarray}
V_{NN}^C (r) = - \frac{32 \pi}{3 m_\pi^4}\int \frac{d^3 q}{(2 \pi)^3}
e^{i q \cdot x} \left[\sigma_{\pi N} (-q^2) \right]^2 t_{00}
(-q^2)
\end{eqnarray} 
where $\sigma_{\pi N} (s) $ is the $\pi N$ sigma term and $t_{00} (s)
= (e^{2 I \delta_{00}(s)}-1)/( 2 i \sigma(s))$ the $\pi\pi$ scattering
amplitude in the $I=J=0$ channel as a function of the CM energy
$\sqrt{s}$. In the large $N_c$ limit, $ t_{\pi \pi} (s) \sim 1/N_c $
while $\sigma_{\pi N} (s) \sim N_c $ yielding $V_{NN} \sim N_c $ as
expected~\cite{Kaplan:1996rk}. At the sigma pole 
\begin{eqnarray}
\frac{32 \pi}{3 m_\pi^4}\left[\sigma_{\pi N} (s) \right]^2 t_{\pi
\pi}^{II} (s) \to \frac{g_{\sigma NN}^2}{s - (m_\sigma- i
\Gamma_\sigma )^2} \to \frac{g_{\sigma NN}^2}{s - m_\sigma^2}
\end{eqnarray} 
where in the second step we have taken the large $N_c$ limit. This
yields $g_{\sigma \pi\pi} \sim 1/ \sqrt{N_c}$, provided $m_\sigma \sim
N_c^0$ and $\Gamma_\sigma \sim 1 / N_c$, a point disputed in
Ref.~\cite{Pelaez:2003dy} where the IAM method is applied to $\pi\pi$
scattering.  If we use instead the Bethe-Salpeter method to lowest
order~\cite{Nieves:1999bx}, we get a once subtracted dispersion
relation, with an arbitrary constant
\begin{eqnarray}
t_{00}^{-1} (s) - t_{00}^{-1} ( 4 m_\pi^2 ) = v_{00}^{-1} (s) -
v_{00}^{-1} ( 4 m_\pi^2 ) + \frac1{\pi}\int_{4 m_\pi^2}^\infty \, ds' \, 
\sigma(s') \left[\frac1{s-s'}-\frac1{s-4m_\pi^2} \right]
\label{eq:bs}
\end{eqnarray} 
where $v_{00} (s) = (m_\pi^2 -2 s) / (32 \pi^2 f_\pi^2) $ is the tree
level amplitude and $\sigma(s) = (1-4 m_\pi^2 /s )^\frac12 $ the
two-pion phase space. The difference between $t_{00}(4 m_\pi^2)$ and
$v_{00} ( 4 m_\pi^2)$ is higher order in the chiral expansion but
both scale as  $1/N_c$.  We fix the accurately determined
scattering length $ -t_{00} ( 4 m^2) = a_{00} m = 0.220(2)
$~\cite{Nieves:1999zb,Colangelo:2001df}.  For $f_\pi=92.3 {\rm MeV}$,
and $m=139.6 {\rm MeV}$ we get the pole at $m_\sigma- i \Gamma_\sigma
/2 = 467 - i 192 {\rm MeV} $
although $\delta_{00} = 50^o$ at $E_{\pi\pi}=500 {\rm MeV}$ overshoots
the Roy analysis value $\sim 35(5)^0$~\cite{Colangelo:2001df} mainly
because higher order chiral corrections~\cite{Nieves:1999bx} and
possibly subthreshold $K \bar K$ effects~\cite{Oller:1997ng}, have been
omitted.  Scaling according to large $N_c$ counting $a_{00}
\to \sqrt{3/N_c} a_{00} $ and $f_\pi \to \sqrt{N_c/3} f_\pi$ the
unitarity integral in Eq.~(\ref{eq:bs}) can be neglected and the pole satisfies
\begin{eqnarray}
- (a_{00} m_\pi)^{-1}= v^{-1}_{00} (m_\sigma^2) - v^{-1}_{00} ( 4
m_\pi^2 )
\end{eqnarray} 
The limit is smooth, and while $\Gamma_\sigma \to 0$ we get $m_\sigma
\to 506.8 {\rm MeV}$, closer to the $NN$-scalar. 
On view of this agreement it is tempting to think that perhaps the
$\sigma$ proposed by Johnson and Teller in 1955 might correspond to
the $\sigma$ determined by Caprini, Colangelo and Leutwyler in 2006
{\it in the large} $N_c$ limit. It remains to be seen if higher order
chiral and $1/N_c$ corrections both for $NN$ as well as for $\pi\pi$
support this view.

\bigskip
Supported by  Spanish DGI and FEDER
funds with grant FIS2005-00810, Junta de Andaluc{\'\i}a grant
FQM225-05, and EU Integrated Infrastructure Initiative Hadron Physics
Project contract RII3-CT-2004-506078.


\bibliographystyle{aipprocl} 


\end{document}